**The Coronal Global Evolutionary Model: Using HMI Vector Magnetogram and Doppler Data to Model the Buildup of Free Magnetic Energy in the Solar Corona**

G. H. Fisher, W. P. Abbett, D. J. Bercik, M. D. Kazachenko, B. J. Lynch, and B. T. Welsch
(Space Sciences Laboratory, University of California, Berkeley)

J. T. Hoeksema, K. Hayashi, Y. Liu, A. A. Norton, A. Sainz Dalda, and X. Sun,
(Stanford University)

M. L. DeRosa, and M. C. M. Cheung
(Lockheed Martin Solar and Astrophysics Laboratory)

The most violent space weather events (eruptive solar flares and coronal mass ejections) are driven by the release of free magnetic energy stored in the solar corona. Energy can build up on timescales of hours to days, and then may be suddenly released in the form of a magnetic eruption, which then propagates through interplanetary space, possibly impacting the Earth's space environment. *Can we use the observed evolution of the magnetic and velocity fields in the solar photosphere to model the evolution of the overlying solar coronal field, including the storage and release of magnetic energy in such eruptions?* The objective of the Coronal Global Evolutionary Model, funded by the NASA/NSF Space Weather Modeling program, is to develop and evaluate such a model for the evolution of the coronal magnetic field. The evolving coronal magnetic field can then be used as a starting point for magnetohydrodynamic (MHD) models of the corona, which can then be used to drive models of heliospheric evolution and predictions of magnetic field and plasma density conditions at 1AU.

The Coronal Global Evolutionary Model (CGEM) effort can be broadly described as follows: (1) Develop enhanced processing and analysis of vector magnetic field and Doppler velocity data taken with the Helioseismic and Magnetic Imager (HMI) instrument on NASA's Solar Dynamics Observatory (SDO) Mission; (2) Use these processed data to derive the electric field in the solar photosphere; (3) Using the photospheric magnetic and electric field vectors, evolve the overlying coronal magnetic field, using a magnetofrictional model (MF) for the quasi-static magnetic evolution as energy builds up; and (4) for magnetic field configurations that become unstable, use MHD models to evolve the solar corona, using output from the MF model as an initial condition. The MF model replaces the inertial forces that are present in MHD with a "frictional" force that allows one to efficiently model the quasi-static build-up of magnetic energy in the coronal magnetic field (Cheung & DeRosa 2012). Step (4) of our effort was pioneered by Pagano et al. (2013), using MF models (Mackay and van Ballegooijen 2006; Yeates, 2014) as a starting point for MHD simulation. Working prototype versions of the above components of CGEM exist now, but we don't expect the final, production version of these parts of CGEM to be complete until the 2016-2018 time frame. Annual reports on the detailed progress of CGEM can be viewed on the website http://cgem.stanford.edu/.

For regions of the Sun for which no magnetic field data are available, a flux transport model will be developed and used to estimate the evolution of the magnetic and electric fields for the global version of the magnetic field model discussed below.

We plan for the CGEM models to include both a lower-resolution global configuration, and a higher-resolution model describing active region (AR) temporal and spatial scales. Both models will use spherical coordinates. We have successfully developed a Cartesian version of the AR scale electric-field inversion model, and the corresponding Cartesian MF model, that we are using as a prototype for the spherical models.

The primary innovation of the CGEM data-driven models, as compared to the current generation of potential-field models that are constrained match the surface magnetic field, is the ability of the model corona to support electric currents generated inductively by electric fields, thereby allowing for a consistent energy buildup process in the corona. While several current global MHD models of the corona also contain currents (e.g. Mikic et al 1999; Toth et al 2005; DeVore and Antiochos 2008) they aren't data-driven in the above sense -- the first two of these are data-constrained models (the magnetic field at the photosphere is constrained to instantaneously match the photospheric data), and the latter is a data-inspired model, in which the boundary conditions are artificial, but inspired by observed magnetic field evolution.

**Enhanced HMI Data Processing**
The HMI instrument on SDO (Schou et al., 2012) and the associated pipeline for processing this data (Hoeksema et al., 2014) produces vector magnetic field maps of the entire visible disk of the Sun every 12 minutes. The observed time evolution of the magnetic vector, as well as the Doppler velocity, allows one in principle to derive the electric field through an inversion of Faraday's Law and other constraints described in further detail below. In practice, before this computation can be made reliably, two additional steps are required: (1) the 180-degree disambiguation of the transverse magnetic field component determined in the pipeline must be regularized in time to ensure the evolution is physically reasonable, and (2) an absolute calibration of the Doppler velocity zero point must be made to ensure that the electric field near magnetic polarity inversion lines is consistent with the evolution of nearby fluxes of the normal component of the magnetic field. These procedures are described in Welsch et al. (2013).

The vector magnetic field is observed by HMI to change during eruptive solar flares (e.g. Sun et al., 2012). That information enables the computation of the change in the vector Lorentz force integrated over the volume of the outer solar atmosphere above a flaring AR. Figure 1 shows the three-dimensional nature of the magnetic field of AR 11158 during a snapshot in time on February 14, 2011. The change in the Lorentz force integral can be used to estimate the impulse provided by magnetic forces to the erupting magnetized plasma (Fisher et al., 2012a). Knowing the magnetic impulse allows one to predict the momentum and place limits on the mass of coronal mass ejections using only the photospheric magnetic field data. The CGEM project has implemented the calculation of Lorentz forces in HMI AR patches (Bobra et al. 2014) as a routine data product that is now available at the Solar Dynamics Observatory Joint Science Operations Center (Bogart 2007). The details of the cgem.lorentz data product are described in Sun et al. (2014).

**Finding Electric Fields from Vector Magnetic Field Evolution and Doppler Shifts**
The time evolution of the vector magnetic field is related to the electric field through Faraday's law: $\frac{\partial \boldsymbol{B}}{\partial t} = -\nabla \times c\boldsymbol{E}$. If one can measure $\frac{\partial \boldsymbol{B}}{\partial t}$ from HMI temporal cadences, can one invert this equation and solve for **E**?  The answer is yes, but the solutions are not unique unless additional information is included (Fisher et al. 2010; Fisher et al. 2012b).  In Kazachenko et al. (2014), we show how the inclusion of additional data, such as that from Doppler line-of-sight velocity measurements taken with HMI, and horizontal flows determined from Local Correlation Tracking, can result in accurate solutions in a test case for which the true electric field is known.  Once the electric fields are determined, they can then be used to provide time-dependent photospheric boundary conditions for the MF models of the solar corona.

**Magnetofrictional Models of the Evolving Corona**
The MF code solves the following form of the induction equation (along with auxiliary constraints):

$$\frac{\partial A}{\partial t} = \boldsymbol{u} \times \boldsymbol{B} - \eta \boldsymbol{J} \ (1) ; \boldsymbol{B} = \nabla \times \boldsymbol{A} \ (2); \boldsymbol{J} = \nabla \times \boldsymbol{B} \ (3) ; \boldsymbol{u} = \frac{(\boldsymbol{J} \times \boldsymbol{B})}{\nu} \ (4),$$

where $\nu$ is the magneto-frictional coefficient, and $\eta$ the magnetic diffusivity.

The code is derived from an existing Cartesian MF code (hereafter CMF; see Cheung and DeRosa 2012 for details of the numerical scheme), which has already been parallelized by using domain-decomposition with the Message Passing Interface (MPI) and OpenMP Application Programming Interfaces.  Figure 2 shows a snapshot in the evolution of AR 11158 computed using the CMF version of the code during a 6-day simulation of the evolution of AR 11158.

**MHD Simulations of Unstable Magnetic Configurations in Active Regions**
In order to extend the magnetofrictional models to include the physics of momenta and energy transport during large-scale eruptive events, we developed new numerical methods, and extended the capabilities of the radiative magnetohydrodynamic code RADMHD (Abbett, 2007; Abbett and Fisher 2012).  These improvements allow us to accommodate AR-scale magnetic structures in a computational domain that can encompass the entire upper convection zone-to-corona system. For CGEM, however, we will be using RADMHD to model only the region above the photosphere, taking advantage of RADMHD's ability to accommodate multi-scale dynamics, from ARs to global scales in spherical coordinates.

The development of this new, spherical, radiative-MHD model was funded, in part, by this NASA-NSF Strategic Capability, and the results from it are one of the intended deliverables of the project.  The goal is to use results from the spherical MF model (along with the driving boundary conditions associated with the MF runs that lead to interesting eruptions) to initiate and drive AR-scale, dynamic, data-driven models of a solar eruption over shorter timescales, correctly evaluating the inertial terms missing from the MF model.  The thermodynamics and equation of state in RADMHD will be simplified to ensure that the initial conditions generated by the MF model are compatible with the MHD model.

**Global MHD Models of the Outer Corona**
Finally, the CGEM effort will couple output from either active region-scale simulations or the global version of the MF model to global-scale MHD simulations of the outer corona (Hayashi 2013).

**Summary**

Progress with the CGEM effort is shown in Figures 1 and 2.  Figure 1 shows a snapshot in time of the vector magnetic field configuration of AR 11158 measured by HMI on 2015 February 14 at 2200UT.   Red and blue colors indicate the direction of the radial component of magnetic field at the surface, while the arrows indicate both the amplitude and the direction of the magnetic field vector.  Figure 2 shows the magnetic field lines in the model corona above AR 11158 computed with the MF model at 01:42UT on February 15, 2011.  The brightness of the field-lines indicates the electric current density, as described in detail in Cheung and DeRosa (2012).  Briefly, a large number of field-lines were traced from starting points at the bottom boundary. Each position of each field-line was assigned an intensity that was proportional to the mean current density squared along that field-line.  The sum of the intensities along a given line of sight then defines the overall brightness in the image (log-scaled). The left half of the Figure shows a view looking down on the corona from above, while the top right portion of the image shows a side view of the corona seen from the bottom of the page, while the bottom right shows a view from the left side of the page.  In all three elements of the Figure, the location of an erupting flux rope in the simulation is labeled.  The onset of the rising motion of the rope occurs around the time of an observed X2.2 eruptive flare in AR 11158.

The different steps involved in the CGEM simulations require different levels of computing resources.  For example, using our current versions of the data post-processing software, the enhanced HMI data products are produced at a faster cadence than the data are recorded, using a standard desktop workstation.  The electric field solutions for the entire AR11158 six-day sequence can be processed in 1-2 hours on an Apple Macbook Pro.   The MF model is more computationally demanding.  The MF simulation, using a domain of 660x624x256 cells required 107,000 CPU hours to reach the state shown in Figure 2 (4.5 days in solar time after the start of the simulation).  6.2 days of wall clock compute time was required to run the simulation on sixty 12-core Westmere nodes of the NASA Pleiades cluster.  Using more nodes (processors), the same model could be run in near real time.

The CGEM investigators are enthusiastic about collaborating with the other Strategic Capability modeling efforts to use CGEM output into other models.

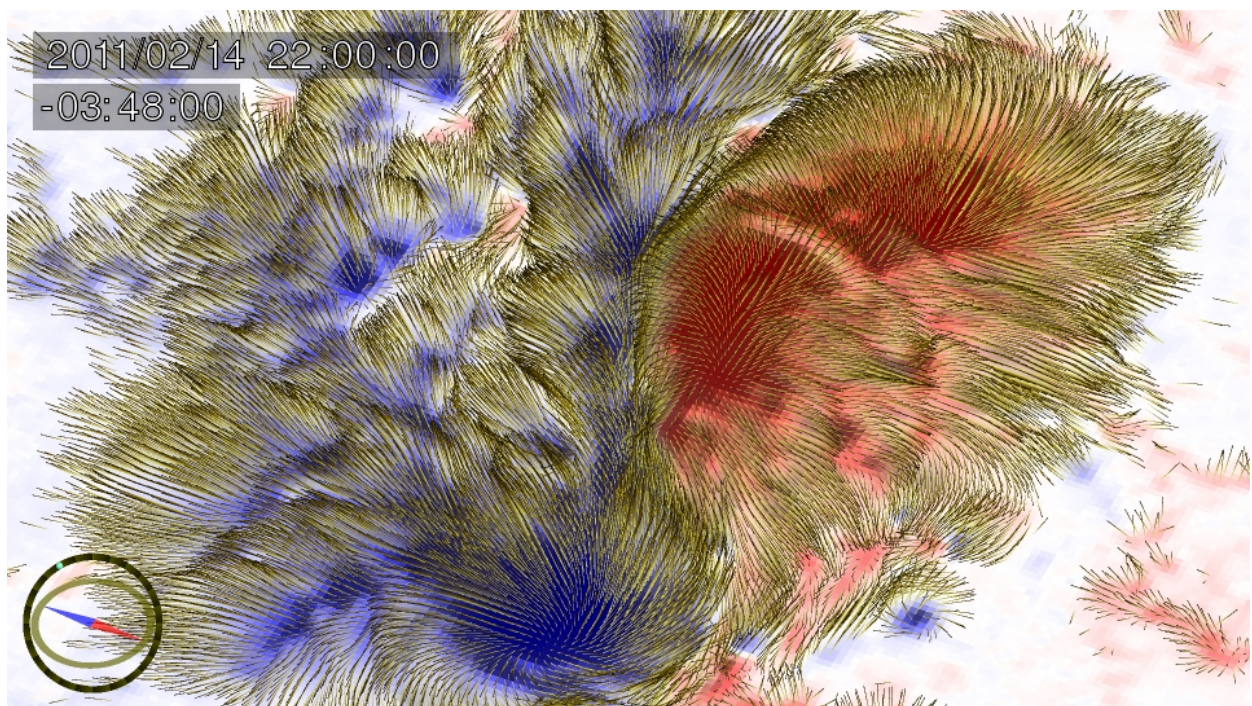

**Figure 1 - Vector magnetic field from HMI measurements of AR 11158. Image is focused along the sheared magnetic neutral line just a few hours before a flux rope originating from this location resulted in an eruptive flare.**

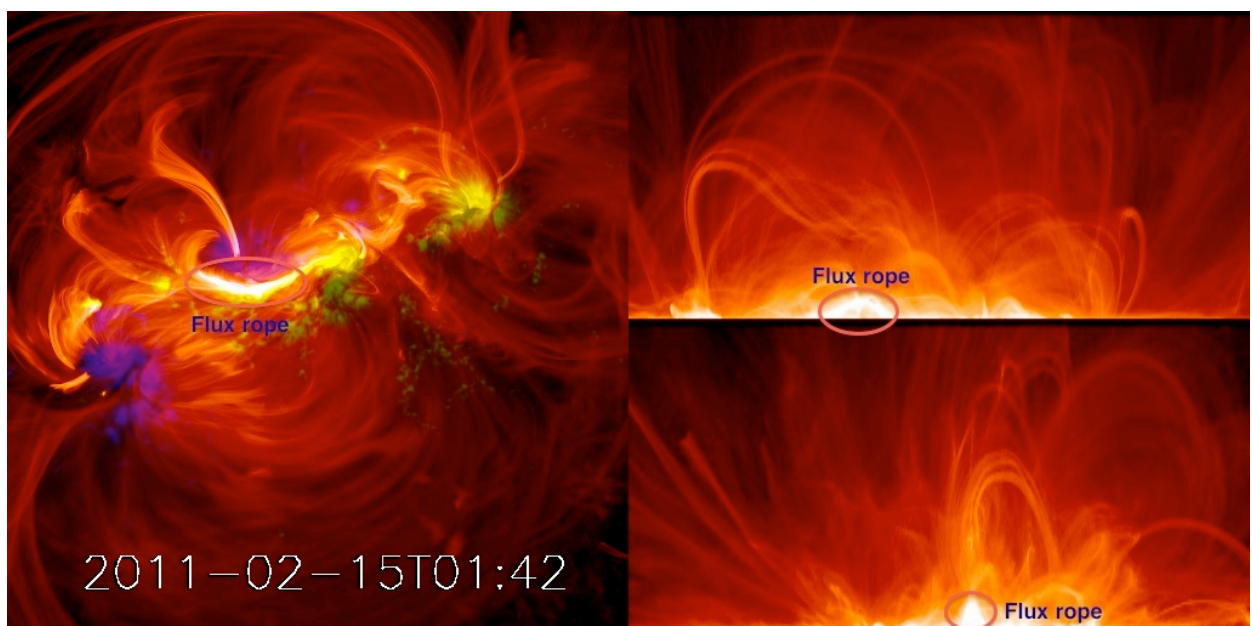

**Figure 2 - Computed evolution of the magnetic field in AR 11152 from the magnetofrictional model, when the erupting flux rope from the model is just starting to lift off. This is also near the time of the onset of the eruptive X2.2 flare that was observed in this active region.**

## Acknowledgements:

This work is supported by NSF and NASA through grants AGS-1321474, NNX13AJ96G, and NNX13AK39G to UC Berkeley, Lockheed Martin, and Stanford University. We thank the US Taxpayers for their support. We thank the anonymous referees for improving the readability and accuracy of this paper. The input data used for CGEM are available from NASA's JSOC website http://jsoc.stanford.edu/, and the output data from CGEM will be made available as a part of the project.

George Fisher is a Senior Space Fellow and Research Physicist at the UC Berkeley Space Sciences Laboratory.

Bill Abbett is a Senior Fellow and Research Physicist at the UC Berkeley Space Sciences Laboratory.

Dave Bercik is a Project Physicist at the UC Berkeley Space Sciences Laboratory.

Maria Kazachenko is an Assistant Research Physicist at the UC Berkeley Space Sciences Laboratory.

Ben Lynch and Brian Welsch are Associate Research Physicists at the UC Berkeley Space Sciences Laboratory.

Todd Hoeksema, Aimee Norton, and Yang Liu are Senior Research Scientists in the Solar Observatories Group in the W.W. Hansen Experimental Physics Laboratory at Stanford University.

Xudong Sun is a Research Associate in the Solar Observatories Group in the W.W. Hansen Experimental Physics Laboratory at Stanford University.

Keiji Hayashi is a Solar Scientist in the Solar Observatories Group in the W.W. Hansen Experimental Physics Laboratory at Stanford University, and is also an Associate Professor at the Solar Terrestrial Environment Laboratory at Nagoya University in Aichi, Japan.

Alberto Sainz Dalda is a Fellow at the Stanford-Lockheed Institute for Space Research.

Marc DeRosa and Mark Cheung are Staff Physicists at the Lockheed Martin Advanced Technology Center's Solar and Astrophysics Laboratory.